\documentclass[12pt]{article}
\usepackage{latexsym}
\usepackage{graphicx}
\usepackage{caption2}
\usepackage{psfrag}
\usepackage{amsmath}
\newcommand{\be}{\begin{equation}}
\newcommand{\ee}{\end{equation}}
\newcommand{\bea}{\begin{eqnarray}}
\newcommand{\eea}{\end{eqnarray}}

\newcommand{\nn}{\nonumber}
\newcommand{\lesssim}{ {\
\lower-1.2pt\vbox{\hbox{\rlap{$<$}\lower5pt\vbox{\hbox{$\sim$}}}}\ } }
\newcommand{\gtrsim}{ {\
\lower-1.2pt\vbox{\hbox{\rlap{$>$}\lower5pt\vbox{\hbox{$\sim$}}}}\ } }

\newcommand{\psibar}{{\overline\psi}}
\newcommand{\ie}{{\it i.e.}}
\newcommand{\eg}{{\it e.g.}}
\newcommand{\cf}{{\it cf.}}
\newcommand{\Mhat}{{\hat M}}
\newcommand{\Fhat}{{\hat F}}
\newcommand{\fhat}{{\hat f}}

\long\def\symbolfootnote[#1]#2{\begingroup%
\def\thefootnote{\fnsymbol{footnote}}\footnote[#1]{#2}\endgroup}

\input epsf

\begin{document}

\begin{titlepage}

\begin{flushright}
\end{flushright}
\vspace*{1.5cm}
\begin{center}

{\Large \bf On the relation between low-energy constants and resonance saturation}\\[2.0cm]

{\bf Maarten Golterman\symbolfootnote[1]{Permanent address:
Department of Physics and Astronomy, San Francisco State University, 1600 Holloway
Ave, San Francisco, CA 94132, USA
} } and {\bf Santiago  Peris}\\[1cm]

Grup de F{\'\i}sica Te{\`o}rica and IFAE\\ Universitat
Aut{\`o}noma de Barcelona, 08193 Barcelona, Spain

\end{center}

\vspace*{1.0cm}

\begin{abstract}
Although there are phenomenological indications that the low-energy constants in the
chiral lagrangian may be understood in terms of a finite number of hadronic
resonances, it remains unclear how this follows from QCD. One of the arguments
usually given is that low-energy constants are associated with chiral symmetry
breaking, while QCD perturbation theory suggests that at high energy chiral symmetry
is unbroken, so that only low-lying resonances contribute to the low-energy
constants. We revisit this argument in the limit of large $N_c$, discussing its
validity in particular for the low-energy constant $L_8$, and conclude that QCD may
be more subtle that what this argument suggests. We illustrate our considerations in
a simple Regge-like model which also applies at finite $N_c$.

\end{abstract}

\end{titlepage}

\setcounter{footnote}{0}

\section{Introduction}

The low-energy physics of the Goldstone bosons associated with the spontaneous
breaking of chiral symmetry is described by the chiral lagrangian, the construction
of which is solely based on the general principles of quantum field theory and the
symmetries of the underlying fundamental theory of the strong interactions, QCD. The
only ``dynamical" information present in the chiral lagrangian connecting it to QCD
is encompassed by the values of the low-energy constants (LECs) that parameterize
the terms appearing in the chiral lagrangian.   At lowest order, there are two LECs,
the pion decay constant $f_\pi$ and the quark condensate
$\langle\psibar\psi\rangle$, whereas at higher orders, a growing number of LECs
appear.  In  the theory with three flavors, there are ten physical LECs at order
$p^4$ \cite{gl}, denoted as $L_i$, $i=1,\dots,10$.

It is thus an interesting question to ask what physics of the strong interaction leads to the
values of the LECs observed in nature.  Since the physical spectrum contains infinite
towers of resonances with given quantum numbers (spin, flavor and parity), it is
natural to think of the values of the order $p^4$ (and higher-order) LECs as resulting
from integrating out these resonances \cite{swisscatalan,MHA}.%
\footnote{The physics of $f_\pi$ and $\langle\psibar\psi\rangle$ is associated with
the breaking of chiral symmetry by the vacuum.}  In particular, sum rules have been
derived relating the order-$p^4$ constants to the masses and decay constants of these
resonances.

Here we revisit in particular the sum rule relating $L_8$ to the scalar and
pseudoscalar resonances with isospin 1, working in the large-$N_c$ and chiral
limits.  The choice of taking $N_c$ large simplifies the discussion because
hadronic resonances are stable in that limit.  Furthermore, it is natural to work in
the chiral limit, since the LECs are independent of the quark masses.   For
comparison, we will also discuss the sum rule relating $L_{10}$ to vector and
axial-vector resonances with isospin 1.

We begin with reviewing the usual type of sum rule relating $L_8$ to resonance
parameters through the difference of scalar and pseudoscalar two-point functions
$\Pi_S(q^2)-\Pi_P(q^2)$, and show that it is in principle ill-defined.   We show
that the problem originates from the fact that each of these two-point functions is
quadratically divergent in QCD. We will then derive a new sum rule, starting from a
subtracted dispersion relation, in Sec.~2.2.   In Sec.~3 we give an example of how
the new sum rule works, using a simple Regge-like model in which the exact answer is
known.  This model allows us also to consider the situation at finite $N_c$. In
Sec.~4 we test the phenomenological approach in which  the difference
$\Pi_S(q^2)-\Pi_P(q^2)$ is saturated by a finite number of resonances. The usual
argument for this approach is that this difference vanishes in perturbation theory,
suggesting that the contribution from resonances higher in the spectrum cancels, at
least approximately. Applying this approach to our model, we find that the effective
parameters (\ie\ masses and coupling constants) which are required to fit
high-energy constraints as well as to yield the correct value for $L_8$ are not the
same as those of the original infinite spectrum. The final section contains our
conclusions.

\section{The large-$N_c$ limit}

\subsection{Review}

In the large-$N_c$ limit, $L_8$ is given by \cite{L8} \be \label{L8}
16B^2L_8=\Delta\Pi(0)=\frac{1}{2}\left(\Pi_S(0)-\Pi'_P(0)\right)\ , \ee in which the
chiral limit is taken, and where we define \be \label{SPfunctions}
\Pi_{S,P}(q^2)=i\int
d^{4}x\;e^{iqx}\;\langle0|T\left\{J_{S,P}(x)J^\dagger_{S,P}(0)\right\}|0\rangle\ ,
\ee with $J_S(x)=\overline{d}(x)u(x)$ and $J_P(x)=\overline{d}(x)i\gamma_5u(x)$. The
order-$p^2$ constant $B$ is equal to $-\langle\psibar\psi\rangle/f_\pi^2$ in the
chiral limit. The prime on $\Pi'_P(0)$ in Eq.~(\ref{L8}) indicates that we omit the
pion pole present in the pseudoscalar two-point function $\Pi_P(q^2)$. Since the
difference $\Pi_S-\Pi_P$ cancels in perturbation theory for vanishing quark masses,
this means that $\Delta\Pi(q^2) $ obeys an unsubtracted dispersion relation in the
chiral limit, \ie\
\begin{equation}\label{ubs}
    \Delta\Pi(q^2)\equiv \frac{1}{2}\left(\Pi_S(q^2)-\Pi'_P(q^2)\right)=
\frac{1}{2\pi}\int_0^\infty dt\;\frac{{\rm
Im}\;\left(\Pi_S(t)-\Pi'_P(t)\right)}{t-q^2-i\epsilon}\ .
\end{equation}
Equation (\ref{L8}) states, then, that $L_8$ becomes a physical quantity at
$N_c=\infty$. In this limit, there is an infinite tower of infinitely narrow
resonances in each channel, and we thus have that \cite{tHw} \bea \label{towers}
\frac{1}{\pi}\;{\rm
Im}\;\Pi_S(t)&=&2\sum_n^\infty F^2_{S}(n)\ \delta(t-M^2_{S}(n))\ ,\nn \\
\frac{1}{\pi}\;{\rm Im}\;\Pi'_P(t)&=&2\sum_n^\infty F^2_{P}(n)\
\delta(t-M^2_{P}(n))\ . \eea Use of Eq.~(\ref{towers}) in Eq.~(\ref{ubs}) leads
formally to \bea \label{SPagain} \Delta\Pi(q^2) &=&\sum_n^\infty
\frac{F^2_{S}(n)}{M^2_{S}(n)-q^2-i\epsilon}- \sum_n^\infty
\frac{F^2_{P}(n)}{M^2_{P}(n)-q^2-i\epsilon}\ . \eea In arriving at this expression
for $\Delta\Pi(q^2)$, we assumed that the integral over $t$ and the sum over $n$ may
be freely interchanged.  Ignoring the question of validity of this assumption for
the time being, one thus arrives at a sum rule for $L_8$: \be \label{L8naive}
16B^2L_8=\sum_n^\infty\frac{F^2_{S}(n)}{M^2_{S}(n)}-\sum_n^\infty\frac{F^2_{P}(n)}{M^2_{P}(n)}\
. \ee As it stands, this sum rule is not well-defined, because both sums appearing
on the right-hand side are quadratically divergent. First, let us give an example.
If we assume Regge-like behavior asymptotically for large $n$, with\footnote{For the
asymptotic proportionality between decay constants and the masses, see
Eq.~(\ref{kappa}) below.} \bea \label{Regge}
M^2_{S,P}(n)&\sim&n\Lambda^2\ ,\ \ \ \ \ n\ {\rm large}\ ,\\
F^2_{S,P}(n)&\sim&\kappa\Lambda^2M^2_{S,P}(n)\ ,\nonumber \eea with $\Lambda$ a
constant of order $1$~GeV, and $\kappa$  is equal to the coefficient of the parton
model logarithm, then Eq.~(\ref{L8naive}) looks like \be \label{L8naiveRegge}
16B^2L_8\sim\kappa\Lambda^2\left(\sum_{n}^\infty 1-\sum_{n}^\infty 1\right)\ , \ee
which is clearly ill-defined. However, as we will now discuss, the problem is model
independent. We will return to the model (\ref{Regge})  in Sec.~3.

In order to proceed in the general case, let us  introduce cutoffs $N_{S,P}$ on the
summation index $n$ in the scalar and pseudoscalar channels, corresponding to
cutoffs $M_{S}(N_S), M_{P}(N_P)$ on the masses, in order to regulate the sums.
Furthermore, we require that both two-point functions match on to QCD perturbation
theory for asymptotically large euclidean $Q^2=-q^2$.  In particular, the leading
large-$Q^2$ behavior is given  by $\kappa Q^2\log{Q^2}$, with $\kappa$ a constant
common to both channels, because the perturbative expressions for scalar and
pseudoscalar two-point functions, valid for asymptotically large $Q^2$, are equal.
It follows that indeed the number of resonances in each channel has to be infinite
\cite{tHw}. Furthermore, writing \be \label{parton} F^2_{S,P}(n)= f^2_{S,P}(n)
M^2_{S,P}(n)\ , \ee each of the sums in Eq.~(\ref{SPagain}) can be written as
(omitting the indices $S$ or $P$) \bea \label{sum}
\sum_n^N\frac{F^2(n)}{M^2(n)+Q^2}&=&\sum_n^N f^2(n)-Q^2\sum_n^N\frac{f^2(n)}{M^2(n)+Q^2}\\
&&\!\!\!\!\!\!\!\!\!\!\!\!\!\!\!\!\!\!\!\sim\int^{M^2(N)}\!\!\!\!
dM^2\left(\frac{dn}{dM^2}f^2(n)\right) -Q^2\int^{M^2(N)}
\!\!\!\!dM^2\left(\frac{dn}{dM^2}f^2(n)\right)\frac{1}{M^2+Q^2}\ .\nonumber \eea If
$(dn/dM^2)f^2(n)$ goes like a constant for large $M^2$, the second integral will
behave like $\log{\left(M^2(N)/Q^2\right)}$ for $M^2(N)\gg Q^2$, thus yielding the
desired parton-model logarithm. As a consequence, the first integral in
Eq.~(\ref{sum}) is quadratically divergent. Note  that $M^2(N)\gg Q^2$ should hold
for any $Q^2$, and that we thus need to require that $M^2(N)\to\infty$ when
$N\to\infty$. It follows that the constant $\kappa$ can be expressed as \be
\label{kappa} \kappa\equiv
\lim_{n\to\infty}\frac{dn}{dM^2_S}f^2_{S}(n)=\lim_{n\to\infty}\frac{dn}{dM^2_P}f^2_{P}(n)\
. \ee Furthermore, it also follows that $M^2_S(N_S)/M^2_{P}(N_P)\to 1$, to avoid a
term linear in $Q^2$ in $\Pi_{S}(q^2)-\Pi_{P}(q^2)$, for $q^2=-Q^2$.

In fact, we may choose to introduce a common cutoff $\Lambda^2_{co}$ on the $t$
integrals in Eq.~(\ref{SPagain}), which implies cutoffs $N_{S,P}$ on the sums over
$n$ such that $M^2_{S}(N_S),M^2_{P}(N_P) \le\Lambda^2_{co}$, while $M^2_{S}(N_S+1),
M^2_P(N_P+1)>\Lambda^2_{co}$. Using Eq.~(\ref{parton}), our regulated sum rule now
reads \be \label{L8naivereg}
16B^2L_8(\Lambda_{co})=\sum_n^{N_S(\Lambda_{co})}f^2_{S}(n)-
\sum_n^{N_P(\Lambda_{co})}f^2_{P}(n)\ . \ee This is of course finite, by
construction.  However, it does not satisfy the basic field-theoretic requirement
that physics should be independent of the cutoff, \ie\ that the relative change in
$L_8$ due to a change in cutoff goes to zero as we take the cutoff to infinity.
Mathematically, this requirement translates into \be \label{qft}
L_8(\Lambda_{co})=L_8+O\left(\frac{1}{\Lambda_{co}}\right)\ , \ee where $L_8$ is the
value of the LEC in the limit of infinite cutoff.  That Eq.~(\ref{L8naivereg}) does
not in general satisfy this requirement can be seen as follows.  When we increase
$\Lambda_{co}$, both $N_S$ and $N_P$ increase by integer steps when $\Lambda_{co}$
moves past the next resonance mass in either channel. Suppose that we increase
$\Lambda_{co}$ such that $N_S$ changes by $1$ to $N'_S=N_S+1$, while $N'_P=N_P$
stays the same. This leads to a change $\Delta L_8$ in $L_8$ equal to \be
\label{deltaL8} \Delta L_8=\frac{f^2_{S}(n=N_S+1)}{16B^2} \ . \ee We do not at
present have enough information about QCD allowing us to conclude that the
right-hand side of this equation goes to zero for $\Lambda_{co}\to\infty$. In fact,
it is \textit{not} the case for the Regge-like behavior of Eq.~(\ref{Regge}), for
which $f^2_{S}(n)$ goes to a constant for large $n$ (\cf\ Eq.~(\ref{kappa})).
Therefore, we conclude that the sum rule for $L_8$ as given by Eq.~(\ref{L8naive}),
even with the cutoff we introduced to regulate the sums in Eq.~(\ref{L8naive}), is
not well-defined.\footnote{Unless there is an $N$ so that $f_{S}(n)=f_{P}(n)$ for
all $n>N$.}

\subsection{A better sum rule for $L_8$}

Obviously, the invalid step in the derivation of the sum rule (\ref{L8naive}) is the
interchange of the integral over $t$ and the sum over $n$, as this led to an
expression containing the difference of two quadratically divergent sums.  We may
remedy this problem by starting from  a once-subtracted dispersion relation for
$\Delta\Pi(q^2)$ \cite{csr},
\be \label{disp}
\Delta\Pi(q^2=-Q^2)=\Delta\Pi(0)-\frac{Q^2}{\pi}\;\int_0^{\infty}dt\;
\frac{\frac{1}{2}\,{\rm Im}\Big(\Pi_S(t)-\Pi'_P(t)\Big)}{t(t+Q^2)} \ , \ee
where again the pion pole has been omitted.  Substituting the large-$N_c$
expressions Eq.~(\ref{towers}) for the spectral functions, one obtains \be
\label{sminp} \Delta\Pi(q^2=-Q^2) =\Delta\Pi(0)
-Q^2\left(\sum_n^{N_S}\frac{f^2_{S}(n)}{ M^2_{S}(n)+Q^2}
-\sum_n^{N_P}\frac{f^2_{P}(n)}{ M^2_{P}(n)+Q^2}\right)\ , \ee where we used
Eq.~(\ref{parton}), and where $N_S$ and $N_P$ are to be taken to infinity.   Again
we interchanged the integral over $t$ with the sum over $n$, but now each sum in
Eq.~(\ref{sminp}) is only logarithmically divergent: \be \label{manip}
\sum_n^N\frac{f^2(n)}{M^2(n)+Q^2}\sim\int^{M^2(N)}dM^2
\left(\frac{dn}{dM^2}f^2(n)\right)\frac{1}{M^2+Q^2}
\sim\kappa\log{\left(M^2(N)/Q^2\right)}\ , \ee (\cf\ Eq.~(\ref{kappa})), and
therefore a cutoff $\Lambda_{co}$ needs to be introduced such that
$Q^2\ll\Lambda_{co}^2\sim M^2_{S}(N_S)\sim M^2_{P}(N_P)$. The limit of $N_S$ and
$N_P$ to infinity thus has to be taken in a correlated way such that this condition
is fulfilled.   It follows that the difference in Eq.~(\ref{sminp}) is finite and
unambiguous. Using the fact that $\Delta\Pi(Q^2)$ should vanish for asymptotically
large $Q^2$, we arrive at a fully regulated sum rule for $L_8$, valid in large-$N_c$
QCD:\footnote{ The pion pole does not play any role in this sum rule.} \be
\label{sumrule} 16B^2L_8=\Delta\Pi(0)=\lim_{Q^2\to\infty}Q^2\lim_{N_s,N_P\to\infty}
\left(\sum_n^{N_S}\frac{f^2_{S}(n)}{ M^2_{S}(n)+Q^2} -\sum_n^{N_P}\frac{f^2_{P}(n)}{
M^2_{P}(n)+Q^2}\right)\ , \ee where it is understood that the limit
$N_S,N_P\to\infty$ is taken in a correlated way, such that
$M^2_S(N_S)/M^2_P(N_P)\rightarrow 1$.

If, as already discussed in Sec.~2.1, we take the number of resonances in each
channel to be finite, assuming that there are $n_S$ in the scalar channel and $n_P$
in the pseudoscalar channel below a certain scale $s_0$, above which we stipulate
that the scalar and pseudoscalar two-point spectral functions exactly cancel, each
of the sums in Eq.~(\ref{sumrule}) is finite, and we may take the large-$Q^2$ limit
under the sums, obtaining \be \label{MHA}
16B^2L_8=\sum_{n=1}^{n_S}f^2_{S}(n)-\sum_{n=1}^{n_P}f^2_{P}(n)\ , \ee \ie\ a finite
version of Eq.~(\ref{L8naive}). In general, if the difference of the sums in
Eq.~(\ref{sumrule}) would be sufficiently convergent, one would be allowed to take
the large-$Q^2$ limit under the sums.  However, it is known that the coefficients of
the OPE in QCD have anomalous dimensions, implying that logarithmic corrections to
the inverse powers of $Q^2$ appear.  This already implies that expanding in $1/Q^2$
under the sums is in general not allowed. In particular, an exact cancellation of
the spectral functions above some $s_0$ cannot occur.

It is instructive to compare our sum rule for $L_8$ with a similar sum rule for
$L_{10}$, which is related to the difference of the vector and axial-vector
two-point functions $\Pi_V(q^2)$ and $\Pi_A(q^2)$ at $q^2=0$.  The key difference is
that in this case each of these two-point functions is only logarithmically
divergent to begin with, because of gauge invariance. The vector and axial-vector
two-point functions are defined by \be \label{VA} \Pi^{\mu\nu}_{V,A}=i\int
dx\;e^{iqx}\;\langle0|T\left\{J^\mu_{V,A}(x)J^{\dagger\nu}_{V,A}(0)\right\}|0\rangle
=(q_\mu q_\nu-g_{\mu\nu}q^2)\Pi_{V,A}(q^2)\ , \ee with
$J^\mu_V(x)=\overline{d}(x)\gamma_\mu u(x)$ and
$J^\mu_A(x)=\overline{d}(x)\gamma_\mu\gamma_5u(x)$.  Perturbation theory tells us
that these two-point functions should both behave like $\kappa'\log{Q^2}$ for
asymptotically large $Q^2$, with $\kappa'$ another constant fixed by the parton
model logarithm for this case. For large $N_c$, the spectra look like \bea
\label{VAtowers} \frac{1}{\pi}\;{\rm Im}\;\Pi_V(t)&=&2\sum_n^\infty F^2_{V}(n)\
\delta(t-M^2_{V}(n))\ ,\\
\frac{1}{\pi}\;{\rm Im}\;\Pi'_A(t)&=&2\sum_n^\infty F^2_{A}(n)\
\delta(t-M^2_{A}(n))\ ,\nonumber \eea where again the prime indicates that we omit
the pion pole.  A line of reasoning similar to that of Sec.~2.1 shows that \be
\label{VAkappa} \kappa'\equiv
\lim_{n\to\infty}\frac{dn}{dM^2_V}F^2_{V}(n)=\lim_{n\to\infty}\frac{dn}{dM^2_A}F^2_{A}(n)\
. \ee {}From this it follows that both $\Pi_V$ and $\Pi'_A$ are logarithmically
divergent, so that no subtraction analogous to that in Eq.~(\ref{disp}) is needed.
We thus obtain the sum rule \bea \label{L10}
-4L_{10}&=&\frac{1}{2}\left(\Pi_V(0)-\Pi'_A(0)\right)\\
&=&\lim_{N_V,N_A\to\infty}\left(\sum_n^{N_V}\frac{F^2_{V}(n)}{M^2_{V}(n)}-
\sum_n^{N_A}\frac{F^2_{A}(n)}{M^2_{A}(n)}\right)\ ,\nonumber \eea where the limit
$N_V,N_A\to\infty$ is again taken in a correlated way such that $\Lambda^2_{co}\sim
M^2_{V}(N_V)\sim M^2_{A}(N_A)$ \cite{gpes,gpope}. Now, if we increase $\Lambda_{co}$
such that, say, $N_V\to N'_V=N_V+1$ while $N_A\to N'_A=N_A$, $L_{10}$ changes by
$\Delta L_{10}= -F^2_{V}(n=N_V+1)/4M^2_{V}(n=N_V+1)$.   If we assume that
$M^2_{V}(n)$ grows like  $n$ (as in Regge-like behavior), we have that $\Delta
L_{10}\sim 1/N_V\to 0$ for $N_V\to\infty$. This is to be compared with the case of
$L_8$ in Eq.~(\ref{deltaL8}).

\section{A Regge-like model at finite $N_c$}

Clearly, the infinite sums in Eq.~(\ref{towers}) are not well-defined mathematical
expressions.  However, at finite $N_c$ the Dirac delta distributions should become
better behaved, and it is therefore illustrative to consider the case of finite
$N_c$. In order to do this, we will consider a simple Regge-like model. In this
model, each scalar resonance will be represented by a pole of the form
\cite{bsz,cgs} \be \label{model} \frac{2F^2_{S}(n)}{z\Lambda^2+M^2_{S}(n)}\ , \ee
with \be \label{z} z=\left(\frac{-q^2-i\epsilon}{\Lambda^2}\right)^\zeta\ ,\ \ \ \ \
\zeta=1-\frac{a}{\pi N_c}\ , \ee and similar for the pseudoscalar resonances, with
masses and residues given by \bea \label{model1}
M^2_{S,P}(n)&=&m^2_{S,P}+n\Lambda^2\ ,\ \ \ \ \ n=0,1,\dots\ ,\\
F^2_{S,P}(n)&=&\kappa\Lambda^2M^2_{S,P}(n)\ ,\nonumber \eea with $m_{S,P}$ also of
order $1$~GeV (\cf\ Eq.~(\ref{Regge})).  For simplicity, we take the parameters
$\Lambda$ and $a$ the same in both channels.  This choice  satisfies
Eq.~(\ref{kappa}).

Expanding the denominator of Eq.~(\ref{model}) to leading order in $1/N_c$, there
are poles near $q^2=M^2_{S,P}(n)(1-ia/N_c)$.  Our model thus describes resonances
with masses $\sim M_{S,P}(n)$ and decay widths \be \label{decay} \Gamma_{S,P}(n)\sim
a M_{S,P}(n)/N_c\ . \ee In the limit $N_c\to\infty$, all resonances are stable. The
function (\ref{model}) has the correct analytic behavior in the complex $q^2$ plane
\cite{bsz}: it is analytic everywhere on the physical sheet, except for a cut along
the positive real axis, starting at $q^2=0$ \cite{bsz}. This model does not include
all corrections one expects at finite $N_c$; in particular, it does not include the
multi-particle continuum starting at $q^2=0$ due to coupling of the sources
$J_{S,P}$ to pions, which would make $L_8$ run. However, as we will see below, the
included finite-$N_c$ behavior already leads to an interesting observation.

The scalar spectral function for our model is given by (for simplicity, we set the
scale $\Lambda$ equal to one in the rest of this section)\footnote{Note that we
start our sums at $n=0$, whereas in Sec.~3 of Ref.~\cite{csr} sums start at $n=1$.}
\be \label{Im} \frac{1}{\pi}\;{\rm Im}\;\Pi_S(t)=\frac{1}{\pi}\sum_{n=0}^\infty
\frac{2F^2_{S}(n)\
\left(t^\zeta\sin(\zeta\pi)+\epsilon\right)}{\left(t^\zeta\cos(\zeta\pi)+M^2_{S}(n)\right)^2
+\left(t^\zeta\sin(\zeta\pi)+\epsilon\right)^2}\ , \ee which for $N_c\to\infty$,
\ie\ $\zeta\to 1$, reproduces Eq.~(\ref{towers}).  The expression for Im~$\Pi'_P(t)$
is similar. These spectral functions are logarithmically divergent with $n$, and we
thus introduce a cutoff $N\equiv N_S=N_P+c$ with $c$ a finite arbitrary constant, so
that $M^2_{S}(N_S)/M^2_{P}(N_P)\to 1$ for $N\to\infty$.\footnote{We may choose
$N_S=N_P+c$ with $c$ an arbitrary finite constant because nothing will depend on $c$
in the limit of infinite cutoff \cite{gpope}.}  The difference between the scalar
and pseudoscalar spectral functions is well defined, and we have that, starting from
the dispersion relation in Eq.~(\ref{ubs}),  \be \label{modelsminp}
\Delta\Pi(0)=\frac{\kappa}{\pi}\int_0^\infty\frac{dt}{t}\sum_{n=0}^\infty\left(
\frac{(n+m_S^2)t^\zeta\sin(\zeta\pi)}
{\left(t^\zeta\cos(\zeta\pi)+n+m_S^2\right)^2+\left(t^\zeta\sin(\zeta\pi)\right)^2}
-(S\to P)\right)\ , \ee where we set $\epsilon=0$, which we are allowed to do as
long as we keep $N_c$ finite.  We may also send $N_c$ to infinity, but in that case
we need to keep an infinitesimal $\epsilon$ in order to reproduce the Dirac
$\delta$-functions of Eq.~(\ref{towers}).

Substituting $y=t^\zeta$ and performing the sum over $n$, we obtain
\bea
\label{yint}
16B^2L_8&=&\Delta\Pi(0)\nn\\
&=&\frac{\kappa}{2i\pi\zeta}\int_0^\infty dy\left(e^{-i\pi\zeta}\psi(ye^{-i\pi\zeta}+m_S^2)
-e^{i\pi\zeta}\psi(ye^{i\pi\zeta}+m_S^2)-(S\to P)\right)\nonumber\\
&=&\frac{\kappa}{2i\pi\zeta}\Bigl[\log\Gamma(ye^{-i\pi\zeta}+m_S^2)-\log\Gamma(ye^{i\pi\zeta}+m_S^2)
-(S\to P)\Bigr]_0^\infty\nonumber\\
&=&\kappa\left(m_P^2-m_S^2\right)\ , \eea where \be \label{psi}
\psi(z)=\frac{d}{dz}\log\Gamma(z) \ee is the digamma function.\footnote{Our explicit
calculation refutes the claim of Ref.~\cite{ae} that $L_8$ vanishes in this type of
model.} Interestingly, in this simple model, the result is {\em independent} of
$N_c$, and it is thus also valid for $N_c\to\infty$.\footnote{We would obtain the
same result setting $N_c=\infty$ from the outset, but taking $\epsilon\to 0$ at the
end of the calculation.}
 The technical reason for this is that setting
$Q^2=0$ in order to obtain $L_8$ removes the dependence on $N_c$, which only appears
through the exponent $\zeta$. We note that the new sum rule Eq.~(\ref{sumrule}) does
apply to our model at finite $N_c$ as well, if one simply replaces $Q^2$ by
$z\Lambda^2$ with $z$ as defined in Eq.~(\ref{z}), and then takes the limit
$z\to\infty$ in Eq.~(\ref{sumrule}).

Let us comment on this result.  First, if we calculate $L_8$ for this simple model
using our new sum rule, Eq.~(\ref{sumrule}), we obtain exactly the same
result.\footnote{ This calculation was done in Sec.~4 of Ref.~\cite{csr}.}  Second,
a key point in the calculation of Eq.~(\ref{yint}) is that we first performed the
sum over $n$, and after that the integral over $t$. These operations do not commute:
if we would interchange the sum and the integral, the result would be different. If
we substitute $y=t^\zeta$ in Eq.~(\ref{modelsminp}) and integrate term by term over
$y$, we obtain for each term \be \label{wrong} \frac{\kappa}{\pi\zeta}\int_0^\infty
dy \frac{(n+m_S^2)\sin{(\pi\zeta)}}
{\left(y\cos{(\pi\zeta)}+n+m_S^2\right)^2+\left(y\sin{(\pi\zeta)}\right)^2} =
\kappa\ , \ee and, restoring $\Lambda$, one would thus find that \be
\label{wrongrule} \Pi'_{S-P}(0)=\kappa\Lambda^2\left(\sum_{n=0}^N\
1-\sum_{n=0}^{N-c}\ 1\right)=c\kappa\Lambda^2\ . \ee Clearly, the result
(\ref{wrongrule}) does not equal the correct result, Eq.~(\ref{yint}). Moreover, it
depends on the undetermined constant $c$, \ie\ on the details of the regulator. If,
as in Sec.~2.1, we increase $\Lambda_{co}$, $c$ alternates between the values $0$
and $1$ (taking $m^2_S<m^2_P$), and the value of $16B^2\Delta L_8$ defined in
Eq.~(\ref{deltaL8}) between $\kappa\Lambda^2$ and $0$. This simple example
calculation demonstrates precisely what goes wrong in the derivation of the naive
sum rule Eq.~(\ref{L8naive}): as we already noted in Sec.~2.1, that equation is
obtained by first performing the $t$ integrals in Eq.~(\ref{SPagain}) (without
subtraction), and then the sums over $n$, thus leading to the difference of two
quadratically divergent sums.  The derivation of the new sum rule (\ref{sumrule})
avoids this problem by starting from a subtracted dispersion relation,
Eq.~(\ref{disp}).

\begin{figure}
\renewcommand{\captionfont}{\small \it}
\renewcommand{\captionlabelfont}{\small \it}
\centering
\includegraphics[width=5in]{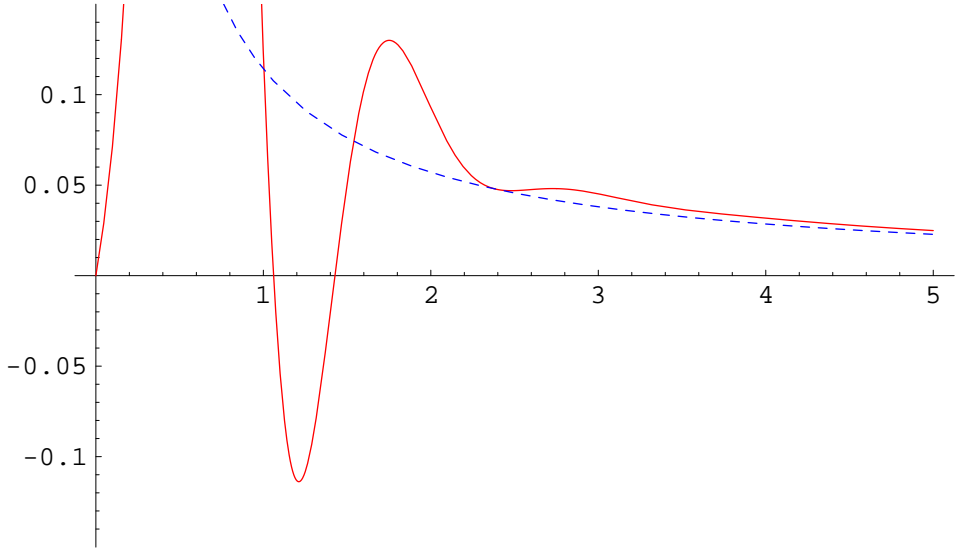}
\caption{Plot of $y$ times the function in Eq. (\ref{yint}) with $m_S^2=(0.8167)^2$
, $m_P^2=(1.083)^2$ and $\zeta=0.82$, as a function of $y$ in units in which
$\Lambda=1$, with arbitrary units on the vertical axis (solid line).  The dashed
 line is the OPE to leading order.}\label{ZZZ}
\end{figure}

Finally, we observe that in our model the result is independent of $N_c$, and thus
it is  equally valid for infinite as well as finite $N_c$. In this respect, it is
interesting to consider the spectral function at finite $N_c$, which we plotted in
Fig.~1.  More precisely, Fig.~1 is a graph of  $y$ times the integrand of
Eq.~(\ref{yint}) as a function of $y$, for a value of $\zeta$ close to unity. We see
that the asymptotic behavior predicted by the OPE sets in already before $y=3$, and
that the resonances at higher values are completely ``washed out" by their growing
decay widths.  In contrast, at infinite $N_c$ the same plot would show an infinite
sequence of Dirac delta functions, \cf\ Eq.~(\ref{towers}). Nevertheless, despite
this very big difference in the behavior of the spectral functions at finite and
infinite $N_c$, the value of $L_8$ is the same in both cases. Even though the
perturbative tail cancels out in the difference between the scalar and pseudoscalar
channels, $L_8$ is not saturated by a finite number of resonances in this model. The
contribution to the integral in Eq.~(\ref{yint}) coming from the region where the
asymptotic behavior already sets in is equally important as that from the region
where resonances are clearly visible.

Our example model can be made more realistic by including a finite number of
resonances in the scalar and pseudoscalar channels below a certain scale $s_0$ which
are not part of the equally-spaced towers of Eq.~(\ref{model}) (so that
$m^2_{S,P}>s_0$). In that case, with $n_{S(P)}$ extra states in the scalar
(pseudoscalar) channel, we would obtain \be \label{L8better}
16B^2L_8=\sum_{n=1}^{n_S}f^2_{S}(n)-\sum_{n=1}^{n_P}f^2_{P}(n)+\kappa (m_P^2-m_S^2)\
. \ee This is to be compared to Eq.~(\ref{MHA}).  The extra term in
Eq.~(\ref{L8better}) originates (in our model)  from the infinite towers present
above $s_0$.  While we do not know what this extra term will look like in the case
of QCD, we do know that at large $N_c$ there is an infinite number of resonances in
each channel.  Therefore, the omission of a term like this in Eq.~(\ref{MHA}) would
represent an unknown systematic error in the application of this sum rule to QCD.

While we certainly do not expect our model to be very realistic, it turns out to be
interesting to apply it to a numerical example, already in the large-$N_c$ limit. In
order to do this, we first want to eliminate the unknown ratio $\kappa/B^2$ from
Eq.~(\ref{yint}). Requiring absence of a $1/Q^2$ term in the OPE expression for
$\Delta\Pi(Q^2)$ in this model, one obtains the additional relation
\cite{csr}\footnote{Here $m_{S,P}$ correspond to the lowest masses in the towers,
whereas in Ref.~\cite{csr} their values were expressed as
$\sqrt{m^2_{S,P}+\Lambda^2}$.} \be \label{ope1}
2B^2f^2=\kappa\left(m^2_S+m^2_P-\Lambda^2\right)\left(m^2_P-m^2_S\right)\ , \ee
where $f=f_\pi$ in the chiral limit.  Combining this with Eq.~(\ref{yint}), we find
\be \label{L8model} L_8=\frac{1}{8}\;\frac{f^2}{m^2_S+m^2_P-\Lambda^2}\ . \ee Taking
$m_S=0.98$~GeV, $m_P=1.3$~GeV, $\Lambda=1.2$~GeV (determined from the $a_0(1450)$
and $\pi(1800)$) and $f=87$~MeV, we find $L_8=8\times 10^{-4}$ to be compared with,
\eg,  $L_8(M_{\rho})=9\times 10^{-4}$ \cite{pich}.

Let us again contrast the case of $L_8$ with that of $L_{10}$, as we already did at
the end of Sec.~2.  In a model with  an equally-spaced tower of resonances in each
channel, one finds for $L_{10}$ \cite{gpes} \be \label{L10es} -4L_{10}= \kappa'
\sum_{n=0}^\infty\frac{(m^2_A-m^2_V)\Lambda^2}{(m^2_V+n\Lambda^2)
(m^2_A+n\Lambda^2)}\ , \ee in which $m_{V,A}$ are the lowest masses in the two
towers.  The key difference is that in this case the contribution from the towers is
not only proportional to the difference $m^2_A-m^2_V$, analogous to the case of
$L_8$, but also suppressed by inverse powers of the product of these resonance
masses. Taking the contribution from the first few resonances in the tower may
already be a good approximation. For instance, with $\kappa'= 1/8\pi^2$ and taking
$m_V=0.77$~GeV, $m_A=1.25$~GeV and $\Lambda=1.3$~GeV gives $L_{10}= -5.6\times
10^{-3}$, to be compared with the full sum in Eq.~(\ref{L10es}) which gives
$L_{10}=-7.1 \times 10^{-3}$. The experimental estimate is $L_{10}(M_{\rho})=-5.5
\times 10^{-3}$ \cite{pich}.

\section{Saturation with a finite number of resonances}

It is often assumed that each spectral function may be approximated by a
\textit{finite} sum over resonances up to a certain scale $s_0$, and a
``perturbative continuum" at values of $t>s_0$.   The perturbative continuum then
cancels between scalar and pseudoscalar channels, leaving the difference between two
finite sums in Eq.~(\ref{MHA}). In these sums, the decay constants and masses of the
resonances are to be considered effective parameters to be fixed by requiring
consistency with the low- and high-momentum expansion of the relevant Green's
function, as obtained from the chiral and operator product expansions, respectively
\cite{swisscatalan}. In the large-$N_c$ limit, this phenomenological \textit{ansatz}
is a rational approximation \cite{Baker} to QCD Green's functions, known as the
Minimal Hadronic Approximation (MHA) \cite{MHA}. It encompasses vector meson
dominance as a particular case.

One may also attempt to use only  the high-energy input from the OPE to predict the
low-energy constants, which parameterize the low-momentum expansion. Although
phenomenologically this assumption seems to work \cite{pich}, it remains unclear how
this follows from QCD. It is also important to be able to control the systematic
error associated with this approximation since, in practice, it is very difficult to
go beyond the inclusion of one resonance per channel. Furthermore, there are often
not enough equations to determine the decay constants and the masses, so that the
effective masses are actually guessed from the position of the physical masses.

In order to gain some insight into these questions a model like that in
Eqs.~(\ref{model},\ref{model1}) is useful because it allows for a comparison of the
phenomenological \textit{ansatz} with the exact large-$N_c$ results produced by the
model. Therefore let us consider the \textit{ansatz}  \be \label{ansatz}
\frac{1}{2}\left(\Pi_S(q^2)-\Pi_P(q^2)\right)_{MHA}
=\frac{\Fhat_S^2}{Q^2+\Mhat_S^2}-\frac{\Fhat_P^2}{Q^2+\Mhat_P^2}
-\frac{B^2f^2}{Q^2}\ , \ee where $\Mhat_{S,P}$ and $\Fhat_{S,P}$ are parameters to
be determined.  The idea is to use short-distance (OPE) constraints to fix these
parameters, and then use their values to predict $L_8$ from \be \label{L8frs}
16B^2L_8=\frac{\Fhat_S^2}{\Mhat_S^2}-\frac{\Fhat_P^2}{\Mhat_P^2}\ . \ee  Writing \be
\label{Ftof} \Fhat^2_{S,P}=\fhat^2_{S,P}\Mhat^2_{S,P}\ , \ee the leading OPE
($1/Q^2$) constraint is \cite{csr} \be \label{ope}
\fhat^2_S\Mhat^2_S-\fhat^2_P\Mhat^2_P=\frac{\kappa}{2} (m_P^2-m_S^2)
(m_S^2+m_P^2-\Lambda^2)\ . \ee With values of $m_{S,P}$ and $\Lambda$ close enough
to 1~GeV, and assuming that $m_P> m_S$, as phenomenology suggests, the right-hand
side of this equation is positive. In fact, this is what happens for the values
chosen after Eq. (\ref{L8model}).

While this one equation is not sufficient to determine the MHA parameters, an
interesting observation can already be made: MHA does not coincide with the physical
masses and decay constants. Suppose that we look for solutions in which the decay
constants take their actual values. In our model, that means
$\fhat^2_S=\fhat^2_P=\kappa\Lambda^2$. One immediately concludes that the ordering
of the mass parameters comes out reversed, \ie\  $\Mhat_S > \Mhat_P$ while the true
ordering is $m_P> m_S$. The mathematical origin of this difference in sign comes
from the fact that the model contains an infinite number of resonances.

To investigate MHA in more detail, let us impose that it gives us the right value of
$L_8$, which adds the relation \be \label{L8eq}
\fhat^2_S-\fhat^2_P=\kappa(m_P^2-m_S^2)\ . \ee The lowest resonance masses are
relatively well known,\footnote{Assuming that the lowest scalar surviving the
large-$N_c$ limit is the $a_0$ with mass 0.98~GeV.} while the $f$ parameters are
much less well known. If one takes the MHA masses as the physical ones, \ie\
$\Mhat_S=m_S$ and $\Mhat_P=m_P$, one can solve for $\fhat_S$ and $\fhat_P$ to find
\bea \label{fhats}
\fhat^2_S&=&\frac{1}{2}\kappa(m_P^2-m_S^2+\Lambda^2)\ ,\\
\fhat^2_P&=&\frac{1}{2}\kappa(m_S^2-m_P^2+\Lambda^2)\ .\nonumber \eea These
solutions are {\em very far} from their ``real-world" values (which, in our test,
are the model values $\fhat^2_S=\fhat^2_P=\kappa\Lambda^2$). Taking the ratio, and
using for instance the numerical example discussed at the end of Sec.~3, we find \be
\label{fhatratio}
\frac{\fhat^2_P}{\fhat^2_S}=\frac{m_S^2-m_P^2+\Lambda^2}{m_P^2-m_S^2+\Lambda^2}
\simeq \frac{1}{3}\ , \ee instead of 1, which is the actual value. Note that both
the low-energy constraint (\ref{L8eq}) and the high-energy constraint (\ref{ope})
force $\hat f^2_S$ to be different from  $\hat f^2_P$ to the extent that $m_S$ is
not equal to $m_P$. Remarkably, nevertheless, the MHA expression (\ref{ansatz})
approximates the true function $\Pi_S-\Pi_P$ to within a few percent for all
euclidean $Q^2$. Similar properties of the Adler function were studied in more
detail in Ref. \cite{Phily}.

One may also try an even simpler \textit{ansatz}, in which only one scalar resonance
(in addition to the pion) is kept. Solving Eqs.~(\ref{ope}) and (\ref{L8eq}) for
$\fhat^2_S$ and $\Mhat^2_S$, we find, using the same numerical example,
$\Mhat^2_S\simeq 0.63\ m_S^2$ and $\fhat^2_S \simeq 0.51\ \kappa\Lambda^2$. As
before, the MHA values are very different from the real ones, even though the
rational approximation to $\Pi_S-\Pi_P$ works again within a few percent for all
euclidean $Q^2$.

\section{Conclusion}

In this paper, we revisited the connection of the order-$p^4$ LEC $L_8$ to the
scalar and pseudoscalar resonance parameters. In Sec.~2.1 we reviewed the fact that
a sum rule of the form (\ref{L8naive}) is generally ill-defined, and showed  that
this is to be expected on general grounds in QCD.     In Sec.~2.2 we derived a
better sum rule, Eq.~(\ref{sumrule}), and argued why this new sum rule gives a
finite and universal result.  Universality follows because each of the sums in
Eq.~(\ref{sumrule}) is only logaritmically divergent, with the difference being
finite, and thus the precise details of the choice of the cutoffs $N_P$ relative to
$N_S$ do not matter, as required by quantum field theory \cite{gpope}. For instance,
increasing $N_S$ to $N_S+1$ in Eq.~(\ref{sumrule}) does not change $L_8$ in the
limit $N_S\to\infty$.

Of course, the ill-defined sum rule of Eq.~(\ref{L8naive}) itself is never used in
practice, but instead each sum is restricted to a relatively small set of resonances
below a scale $s_0$ with adjustable parameters, as reviewed in Sec.~4.   The
derivation of such  sum rules, in this case Eq.~(\ref{MHA}), assumes that above a
certain scale $s_0$ the spectral functions ${\rm Im}\;\Pi_S(t)/\pi$ and ${\rm
Im}\;\Pi_P(t)/\pi$ are exactly equal.  In contrast, our new sum rule,
Eq.~(\ref{sumrule}) relies only on the fact that $\Pi_S$ and $\Pi_P$ become equal
for asymptotically large $Q^2$ in the euclidean regime.

In Ref.~\cite{sppade} it was shown that, by using the theory of Pade approximants
\cite{Baker}, there is a mathematically well-defined connection between resonance
parameters and LECs in large-$N_c$ QCD. However, this work is not applicable in the
case of $L_8$ because the spectral function defining $\Delta\Pi(q^2)$ is not
positive.

The sum rule (\ref{MHA}) can approximate the sum rule (\ref{sumrule}) if the
contribution from the resonances above a certain $s_0$ is very small. However, there
appears to be no reason for this to be true in QCD, unless some new mechanism plays
a role higher up in the spectrum.  A possibility might be that the resonance
parameters of $S$ and $P$ states pair up at higher scales, aligning sufficiently
fast for higher resonances to almost not contribute to our sum rule,
Eq.~(\ref{sumrule}). This scenario of so-called ``parity doubling" or ``chiral
symmetry restoration" has been the subject of much speculation recently, but little
to nothing is known to date about its validity \cite{jps}. In fact, our simple model
demonstrates that this parity doubling in QCD does not follow  from the cancellation
of the perturbative tail in $\Pi_S-\Pi_P$,  and neither does the saturation of $L_8$
by a finite number of resonances.

Clearly, our new sum rule is less practical if indeed the contribution from
resonances above $s_0$ cannot be ignored.   However, ignoring this contribution
anyway would obviously introduce an unknown systematic error. In Sec.~3 we showed
how the new sum rule works in a simple example, in which an explicit form for the
contribution coming from resonances above $s_0$ can be derived, {\it cf.}
Eq.~(\ref{L8better}). While Sec.~2, and in particular the sum rule of
Eq.~(\ref{sumrule}), apply to QCD in the limit of infinite $N_c$,  our model is
defined for any value of $N_c$.  It is semi-realistic in the sense that it has
consistent analytic behavior in the whole complex $q^2$ plane, while it predicts
widths which are suppressed like $1/N_c$, but which grow with increasing resonance
mass.   In particular, we point to Fig.~1, which shows a qualitatively realistic
spectral function for the two-point function $\Delta\Pi(q^2)$. Interestingly, the
result for $L_8$ in that model, Eq.~(\ref{yint}), is \textit{independent} of $N_c$.
We thus speculate that also at finite $N_c$ our general conclusion, that the region
above $s_0$ cannot be ignored, remains valid.

Phenomenologically, Eq.~(\ref{L8frs}) seems to work quite well \cite{pich}. As we
have argued in Sec.~4 using our Regge-like model, this may be at the expense of
values for the parameters in the rational approximant (\ref{ansatz}), which
substantially deviate from the actual ones representing the lowest resonances, but
which produce a very good approximation to the function $\Pi_S-\Pi_P$ over the
entire euclidean region in $Q^2$. Since we lack precise information on the scalar
and pseudoscalar decay constants in the real world, such deviations might go
unnoticed. We conclude that, in spite of the fact that the phenomenology of the
scalar/pseudoscalar sector looks very reasonable, a deeper understanding of the
relation between $L_8$ and the spectrum in the scalar and pseudoscalar sectors in
large-$N_c$ QCD remains an interesting puzzle.

\section*{Acknowledgements}

We thank Oscar Cat\`{a}, Matthias Jamin, Toni Pich and Eduardo de Rafael  for useful
discussions. SP is supported in part by CICYT-FEDER-FPA2005-02211 and SGR2005-00916,
and MG is supported in part by the Generalitat de Catalunya under the program
PIV1-2005 and by the US Department of Energy.


\begin{thebibliography}{99}

\bibitem{gl}
  J.~Gasser and H.~Leutwyler,
  Nucl.\ Phys.\ B {\bf 250}, 465 (1985).

\bibitem{swisscatalan}
  G.~Ecker, J.~Gasser, A.~Pich and E.~de Rafael,
  Nucl.\ Phys.\ B {\bf 321}, 311 (1989);
     G.~Ecker, J.~Gasser, H.~Leutwyler, A.~Pich and E.~de Rafael,
  Phys.\ Lett.\ B {\bf 223}, 425 (1989);
\bibitem{MHA}
  M.~Knecht and E.~de Rafael,
  Phys.\ Lett.\ B {\bf 424}, 335 (1998)
  [arXiv:hep-ph/9712457];
  S.~Peris, M.~Perrottet and E.~de Rafael,
  JHEP {\bf 9805}, 011 (1998)
  [arXiv:hep-ph/9805442].
See also  S.~Peris,
  arXiv:hep-ph/0204181 and
E.~de Rafael,
  Nucl.\ Phys.\ Proc.\ Suppl.\  {\bf 119}, 71 (2003)
  [arXiv:hep-ph/0210317].


\bibitem{L8}
See for instance
  J.~Bijnens, E.~de Rafael and H.~Zheng,
  Z.\ Phys.\ C {\bf 62}, 437 (1994)
  [arXiv:hep-ph/9306323].

\bibitem{tHw}
  G.~'t Hooft,
  Nucl.\ Phys.\ B {\bf 72}, 461 (1974);
    E.~Witten,
  Nucl.\ Phys.\ B {\bf 160}, 57 (1979).

\bibitem{csr}
  O.~Cat\`a, M.~Golterman and S.~Peris,
 Phys.\ Rev.\ D {\bf 74}, 016001 (2006)
   [arXiv:hep-ph/0602194].


\bibitem{gpes}
  M.~Golterman and S.~Peris,
  JHEP {\bf 0101}, 028 (2001)
  [arXiv:hep-ph/0101098].

\bibitem{gpope}
  M.~Golterman and S.~Peris,
  Phys.\ Rev.\ D {\bf 67}, 096001 (2003)
  [arXiv:hep-ph/0207060].



\bibitem{bsz}
 B.~Blok, M.~A.~Shifman and D.~X.~Zhang,
  Phys.\ Rev.\ D {\bf 57}, 2691 (1998)
  [Erratum-ibid.\ D {\bf 59}, 019901 (1999)]
  [arXiv:hep-ph/9709333];
I.~I.~Y.~Bigi, M.~A.~Shifman, N.~Uraltsev and A.~I.~Vainshtein,
  Phys.\ Rev.\ D {\bf 59}, 054011 (1999)
  [arXiv:hep-ph/9805241].





\bibitem{cgs}
  O.~Cat\`a, M.~Golterman and S.~Peris,
  JHEP {\bf 0508}, 076 (2005)
  [arXiv:hep-ph/0506004].

\bibitem{ae}
  S.~S.~Afonin and D.~Espriu,
  arXiv:hep-ph/0602219.


\bibitem{Phily}
 M.~Golterman, S.~Peris, B.~Phily and E.~De Rafael,
  JHEP {\bf 0201}, 024 (2002)
  [arXiv:hep-ph/0112042].






\bibitem{Baker}
For the mathematical theory of rational approximants see, e.g., G.A. Baker and P.
Graves-Morris, ``Pade Approximants,'' Cambridge Univ. Press 1996.

\bibitem{sppade}
  S.~Peris,
  arXiv:hep-ph/0603190.



\bibitem{jps}
See   R.~L.~Jaffe, D.~Pirjol and A.~Scardicchio,
  arXiv:hep-ph/0602010,
and references therein.

\bibitem{pich}
  A.~Pich,
  Int.\ J.\ Mod.\ Phys.\ A {\bf 20}, 1613 (2005)
  [arXiv:hep-ph/0410322].


\end{thebibliography}
\end{document}